# Extension of indoor mmW link radio coverage in non line-of-sight conditions

Mbissane Dieng[1], Gheorghe Zaharia[1], Ghaïs El Zein[1], Raphaël Gillard[1], and Renaud Loison[1]

[1] Univ Rennes, INSA Rennes CNRS, IETR, UMR 6164, F 35000 Rennes, France
mbissane.dieng@insa-rennes.fr

**Abstract.** In future wireless communication systems, millimeter waves (mmWaves) will play an important role in meeting high data rates. However, due to their short wavelengths, these mmWaves present high propagation losses and are highly attenuated by blocking. In this chapter, we seek to increase the indoor radio coverage at 60 GHz in non line-of-sight (NLOS) environments. Firstly, a metallic passive reflector is used in an L-shaped corridor. Secondly, an array of grooved metallic antennas of size 20 cm × 20 cm (corresponding to 80 grooves) is used in a T-shaped corridor. Next, the study focuses on the blockage losses caused by the human body. The results obtained in these different configurations show that it is possible to use beamforming to exploit a reflected path when the direct path is blocked.

## 1 Introduction

The high-data-rate applications have experienced significant growth thanks to the integration of millimeter-band wireless communications. MmWave frequencies, ranging from 30 to 300 GHz, ensure such data rates and increase network capacity. However, due to their short wavelengths, these millimeter waves have high propagation losses, high attenuation by penetration or obstruction and are very attenuated by blocking, which can be caused by the human body.

To provide solutions to these problems, recent research is focused on the integration of massive MIMO and beamforming techniques. In the same time, the environment could be equipped with passive reflectors and/or arrays of active and reconfigurable surfaces. These objects will be placed at intermediate points capable of improving radio coverage and beamforming [1], [2]. In [3], the propagation channel was investigated in order to increase the received power at 28 GHz in NLOS configuration. Reflectors with different shapes were used: flat squares, cylindrical and spherical. When using flat reflectors, it was observed that their orientation is more important than their size. In [4], a cylindrical reflector was used at 60 GHz in an L-shaped corridor for a Tx-Rx separation of 124 m. In [5], the authors designed and experimentally validated a passive reflector operating in order to increase the received power at 60 GHz in a T-shaped corridor. In [6], the effect of passive reflectors was studied to improve NLOS coverage of a mmWave system based on IEEE 802.11ad. The results show that metal reflectors can reduce the path loss (*PL*) by more than 10 dB in some indoor environments. Moreover, one of the first studies of the human blocking [7] was conducted by measurements at 60 GHz for wireless LANs within an office-like environment. Losses of around 20 dB were noted with omnidirectional antennas and 30 dB when using directional antennas [8], [9]. In this experimental study, we consider the optimal position and orientation of the Rx directive antenna according to the type of reflector and the dimensions of different corridors.

Initially, the objective of this work is to increase the power received at 60 GHz in NLOS scenarios, first using a passive reflector in an L-shaped corridor, then using an array of 80 metallic grooves in a T-shaped corridor. Next, the study focuses on the blocking losses introduced by the human body at 60 GHz.

In order to achieve these objectives, measurements and simulations are carried out in these environments. The results obtained, in terms of path loss or channel impulse response, lead to a comparative analysis with and without the reflectors or with and without a human blocker.

The chapter is organized as follows: Section 2 describes a study of passive reflectors in different indoor environments at 60 GHz, then highlights the improvement in radio coverage and the influence of the reflector arrangement. Section 3 focuses on the study of the impact of blocking the radio link by the human body and the possibility of implementing beamforming as a solution against blocking. Finally, Section 4 summarizes several conclusions.

## 2 Propagation using passive reflectors at 60 GHz

### 2.1 Use of a passive reflector in an L-shaped corridor

In this part, we evaluate by measurements the impact of a metal reflector panel on radio coverage in an indoor NLOS scenario at the central frequency fc = 60 GHz.

#### 2.1.1 Measuring system

The used measurement system (Fig. 1) is based on a vector network analyzer (VNA). N = 401 frequency points were considered over a bandwidth of B = 2 GHz, around the intermediate frequency IF = 3.5 GHz. Therefore, the frequency step is $\Delta f$ = 5 MHz. The transmitted power is Pt = 0 dBm.

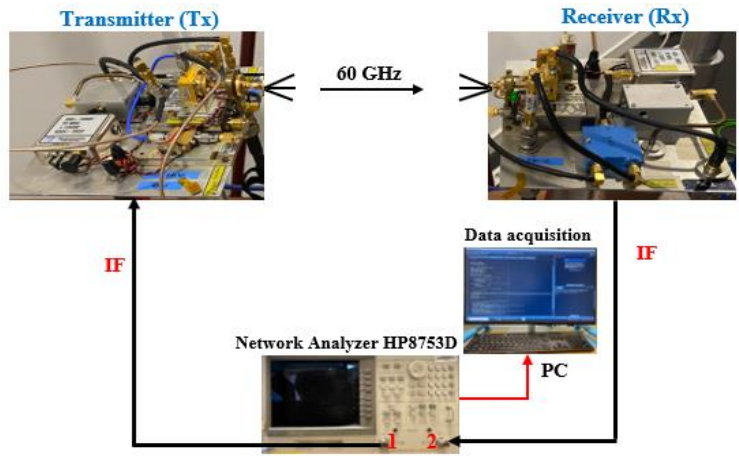

**Fig. 1.** Measurement system.

More details on the measurement system and calculation of the frequency response of the indoor radio channel are given in [10]. Table 1 gives the principal parameters of the used system.

**Table 1**. Parameters of the measurement system.

| | |
|---|---|
| fc (GHz) | 60 |
| B (GHz) | 2 |
| IF (GHz) | 3.5 |
| Δf (MHz) | 5 |
| N | 401 |
| Pt (dBm) | 0 |

For these measurements, two types of antennas were used (Fig. 2).

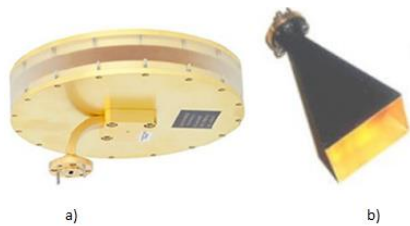

**Fig. 2.** 60 GHz antennas a) omnidirectional b) horn.

An omnidirectional antenna, with a gain of 2 dBi in azimuth and an opening at -3 dB of 30° in elevation was used on the Tx side. The Rx side used a horn antenna, with a gain of 22.5 dBi and a beamwidth at -3 dB of 13° in azimuth and 10° in elevation.

### 2.1.2 Environment and measurement scenario

As shown in Fig. 3, the measurement environment is an L-shaped corridor consisting of two parts A and B whose dimensions are $3.69 \times 2 \times 3$ m$^3$ and $4.7 \times 1.62 \times 3$ m$^3$, respectively [10].

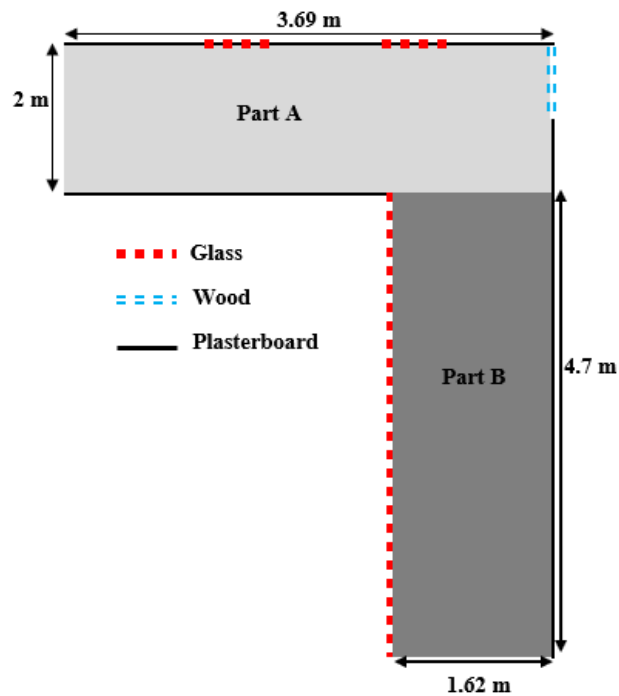

**Fig. 3.** The L-shaped corridor.

The considered corridor has the walls made of plasterboard. The environment was maintained time-invariant, i.e. without any movement. Fig. 4 describes the considered environment. A metal plate (AB), with a length of 98.2 cm and a width of 59.5 cm was used. The panel was placed firstly on a table in vertical position. In this case, the center M of the panel was at a height of $h = 1.37$ m from the ground. This height has been set for Tx and Rx antennas. Moreover, the Rx horn antenna, fixed at a distance $L_R = 3.69$ m from the end of the corridor, was oriented towards the center M of the reflector panel. The omnidirectional Tx transmitting antenna was placed in 16 positions spaced 25 cm apart. Measurements were repeated with a panel placed in horizontal position by keeping the same height of its center M.

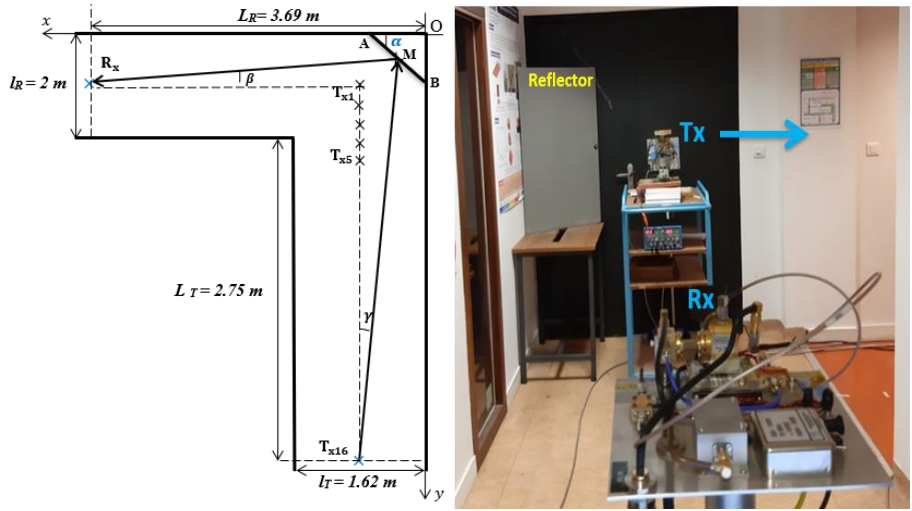

**Fig. 4.** Measurement environment.

Geometric methods are used in [11] to calculate the optimal angles and finally properly orient the panel and the Rx antenna.

### 2.1.3 Measurement and simulation results

For these measurements, we have eliminated the frequency response of the cables by calibration. Then, the frequency response of RF blocks was obtained by back-to-back measurement. The RF frequency response and antenna gains were removed from measured data. We thus obtain the frequency response of the propagation channel. Its module was averaged in linear scale over the 401 frequency points in order to obtain the average path loss for a given Tx position. Fig. 5 shows the obtained path loss (*PL*) for all Tx positions.

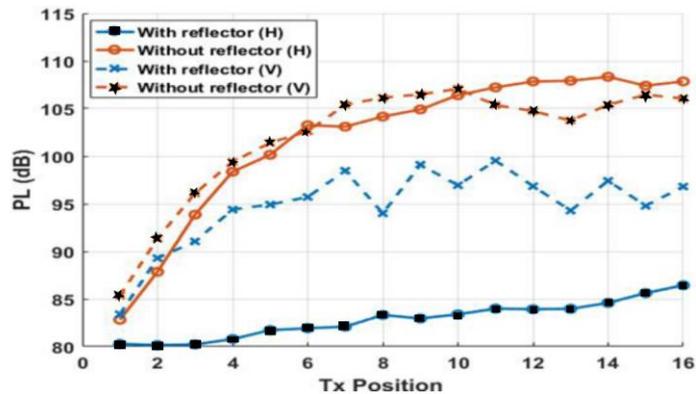

**Fig. 5.** Path loss versus Tx positions.

This figure allows a comparison between the path loss made when the reflector is placed vertically and horizontally, with the same orientation and the same height of its center M.

Overall, we find almost the same results when there is no reflector (Fig. 5) [11]. In this case, there are some differences that may be due to inaccuracies in the positions of the antennas. This difference can also be explained by the fact that the measurements (V) were carried out during one day and the measurements (H) during another day, which means that the back-to-back measurements were perhaps not identical. By using the reflector, it is possible to observe a *PL* decrease up to 24 dB in the horizontal case (H) and a decrease to 12 dB considering the vertical case (V). Thus, in the horizontal case, the reflected energy is greater.

The same scenario was analyzed by simulation carried out with the commercial software WinProp. Fig. 6 gives the *PL* distribution for the position Tx16 with and without the vertical panel.

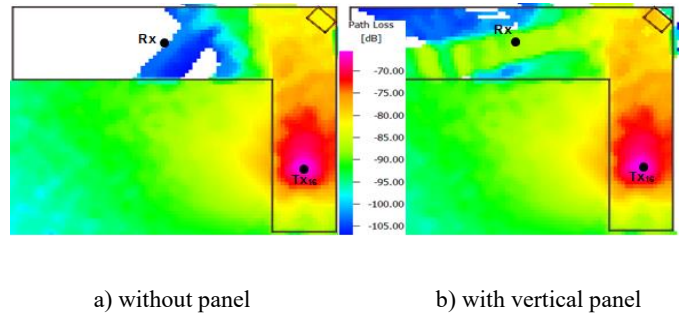

a) without panel　　　　　　　　b) with vertical panel

**Fig. 6.** Simulated path loss in L-shaped corridor.

For the farthest position Tx16 (Fig. 6.a), without the use of the metal panel, the path loss (*PL*) is approximately 105 dB. When the metal panel is used, the *PL* decreases to around 95 dB (Fig. 6.b). These simulation results confirm that the use of the metal panel placed in vertical position can lead to a *PL* reduction of approximately 10 dB, which is close to the result obtained by measurement (Fig. 5). This improves radio coverage in this NLOS configuration.

## 2.2　Use of a passive reflector array in a T-shaped corridor

In this indoor environment, a new measurement campaign was carried out using a groove cell reflector array.

The measurement system used is the same as previously, except that the VNA measures S21 parameter on 801 points over the same frequency band.

An antenna positioning system with a rotation accuracy of 0.02° was used (Fig. 7). Its role is to perform rotations of the Rx antenna in precise steps for a study of the angle of arrival (*AoA*) of the wave in azimuth.

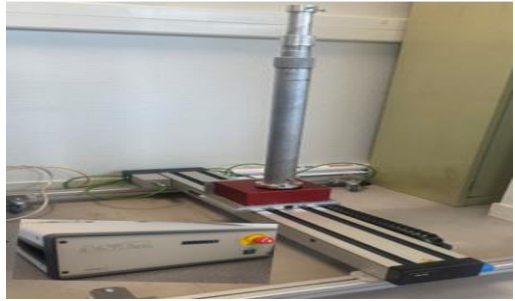

**Fig. 7.** Antenna positioner.

The VNA and the antenna positioner are controlled by a computer via appropriate interfaces, which make it possible to perform antenna rotations using a program developed at IETR INSA and to record the measurement data. Fig. 8 shows the radiation pattern of the horn antennas used for this measurement campaign. They have the same characteristics as the horn antenna used for the previous campaign.

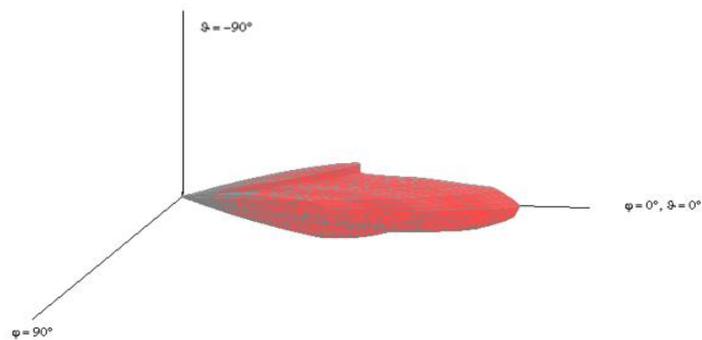

**Fig. 8.** Horn antenna radiation pattern.

### 2.2.1 Environment and measurement scenario

The measurement environment is described in Fig. 9. The walls of the corridors are made of plasterboard. They have a height of 2.7 m. The doors are made of wood. Their dimensions are identical: 2 m height and 0.9 m width.

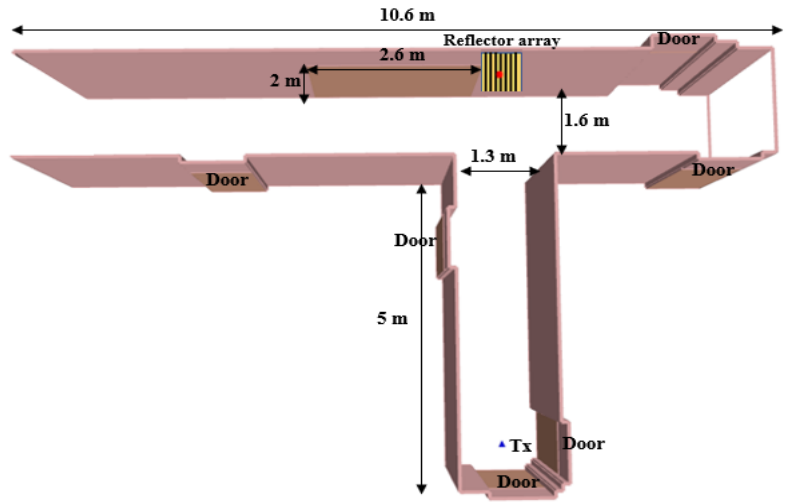

**Fig. 9.** Measurement environment.

As previously mentioned, the two identical horn antennas were used during this measurement campaign. The reflector array was centered on the axis of the short corridor (Fig. 10). The Tx antenna was placed in front of the array at a distance of 5.6 m. The Rx antenna was placed on the axis of the other part of the corridor, perpendicular to the transmitter-reflector axis, in 14 positions spaced 1 m apart. The array is placed 21.5 cm from the wall and its center has been set at the height of the Tx and Rx antennas: 1.7 m.

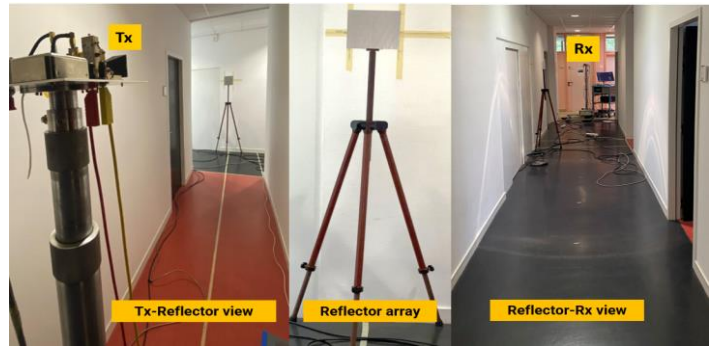

**Fig. 10.** Measurement environment.

For each position of the Rx antenna, a rotation is carried out over 360° in steps of 6°, which allows us to measure the S21 over the 60 angles of arrival (*AoA*) and to identify the direction that corresponds to the maximum received power. During this measurement campaign, two scenarios were studied successively, with and without a reflector array.

### 2.2.2 Design of the reflecting panel

The reflecting panel is designed assuming an impinging plane wave under normal incidence, with *x* polarization. Referring to Fig. 11, the wave vector for the incident wave ($\vec{k}_{inc}$) can be expressed by:

$$\vec{k}_{inc} = -k_0 \vec{e}_z. \tag{1}$$

where $k_0$ is the free space wave number.

The goal is to steer the reflected wave in the ± *x* directions, as shown in Fig. 11, where $\pm\vec{k}_{ref}$ is the wave vector for reflected waves. More precisely, the reflected power is split into two symmetrical beams flowing along the panel surface in the ($\theta$ = 90°, $\phi$ = 0°) and ($\theta$ = 90°, $\phi$ = 180°) directions.

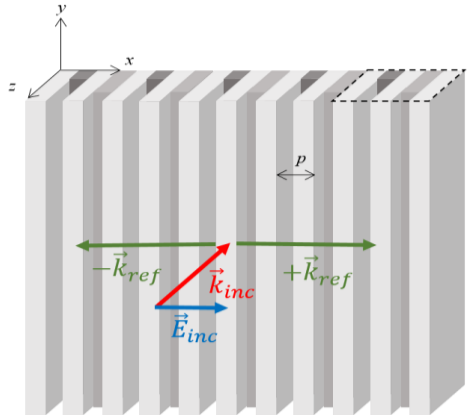

**Fig. 11.** Geometry and principle of the reflecting panel.

The dotted area shows two consecutive cells as detailed in Fig. 12. In order to do so, a phase gradient has to be applied over the reflected panel [12]. In practice, the gradient is produced by discretizing the panel into elementary cells, each of the cells reflecting the incident wave with a desired phase shift. The principle is basically the same as the one used in reflectarray design [13], except that here the illuminating feed is not attached to the reflecting panel but located at $z = +\infty$. Also, as the beams here have to be steered in (*x*, *z*) plane only, no phase variation along the *y* direction is required. Then, the discretization is only done along the *x* axis, with inter-element spacing *p* separating two consecutive cells.

Now, simple antenna array theory shows that the two required reflected beams can be produced by using opposite phase shifts (typically 0° and 180°) on consecutive cells, provided *p* is set to $\lambda_0/2$, $\lambda_0$ being the free space wavelength corresponding to the central frequency ($f_0$ = 60 GHz). Referring to Fig. 12, the reflection coefficients on consecutive cells (for instance $\Gamma_1$ on cell 1 and $\Gamma_2$ on cell 2) must comply with:

$$arg(\Gamma_1) - arg(\Gamma_2) = \pm 180°. \quad (2)$$

$$|\Gamma_1| = |\Gamma_2| = 1. \quad (3)$$

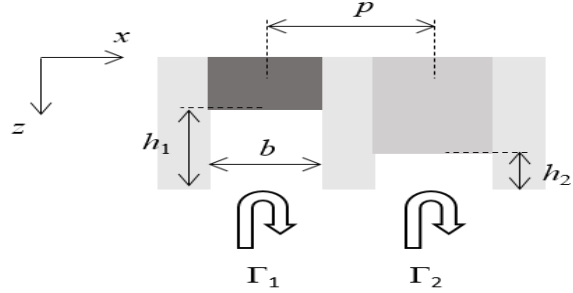

**Fig. 12.** Two consecutive cells and associated reflection coefficients.
(Note: The different grey colors represent the same metal material and are just used for an easier readability of Fig. 11)

The reflecting cells could be printed elements on top of a dielectric layer backed with a ground plane. However, a Metal-Only reflector is preferred here in order to prevent from additional losses in the substrate at 60 GHz and because it naturally provides a full reflection, as required by (3). Moreover, the fabrication is quite straightforward using CNC machining or additive manufacturing and it directly yields a stiff frame that can be handled easily. As shown in Figs. 11 and 12, the reflecting cells consist of vertical grooves (along $y$) in the metal panel whose depths ($h_i$) is optimized to achieve the required phase shifts. The grooves act as Parallel Plate Waveguides (PPW). Most of the incoming wave is converted into the fundamental TEM mode they support. The mode then propagates along $-z$ in the PPW and is reflected after it reaches the short-circuit termination (at distance $h_i$ from PPW input). The phase shift produced by one groove is directly controlled by the distance $h_i$ travelled in the corresponding PPW.

Here, only two different phase shifts and subsequently only two different heights ($h_1$ and $h_2$) are needed. The derivation of the dimensions of the grooves has been explained in [5] and will not be repeated here for the sake of concision. Table 2 summarizes the geometry of the two optimized cells. The panel is then fabricated by alternating cells with heights $h_1$ and $h_2$.

**Table 2.** Geometry of the two different cells used in the panel.

| Cell index ($i$) | $P$ (mm) | $B$ (mm) | $h_i$ (mm) |
|---|---|---|---|
| 1 | 2.5 | 2 | 2.3 |
| 2 | 2.5 | 2 | 0.48 |

The fabricated panel is a 20 cm × 20 cm (40$\lambda_0$ × 40$\lambda_0$) square made from aluminum ($\sigma$ = 37.8×10$^6$ S/m). It involves 80 vertical grooves. A larger panel would intercept more power and produce narrower reflected beams. Indeed, the size can be extended simply with no additional design steps and it has to be chosen depending on the specific application. In addition, for more advanced scenarios, it could be possible to steer also the beam in the (*y*, *z*) plane, as shown for more complex Metal-Only panels in [14] and [15].

The grating consists of 80 groove cells of size 20 cm x 20 cm (Fig. 15) spaced half a wavelength apart. The grating is capable of directing the reflection of the beam towards the lateral directions with a theta angle of approximately ±75° from the normal incidence. The grating has horizontal polarization and the beam width is equal to 10°. Fig. 13 shows a picture of the fabricated panel.

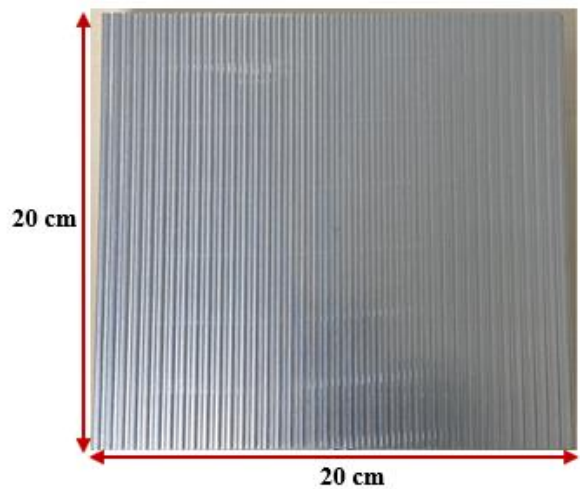

**Fig. 13.** Fabricated metal panel.

### 2.2.3 Measurement results

The results obtained by measurement are shown in Fig. 14, which represents the power received ($P_r$) for each Rx position and for the different angles of arrival (*AoA*), with (red curve) and without reflector (blue curve).

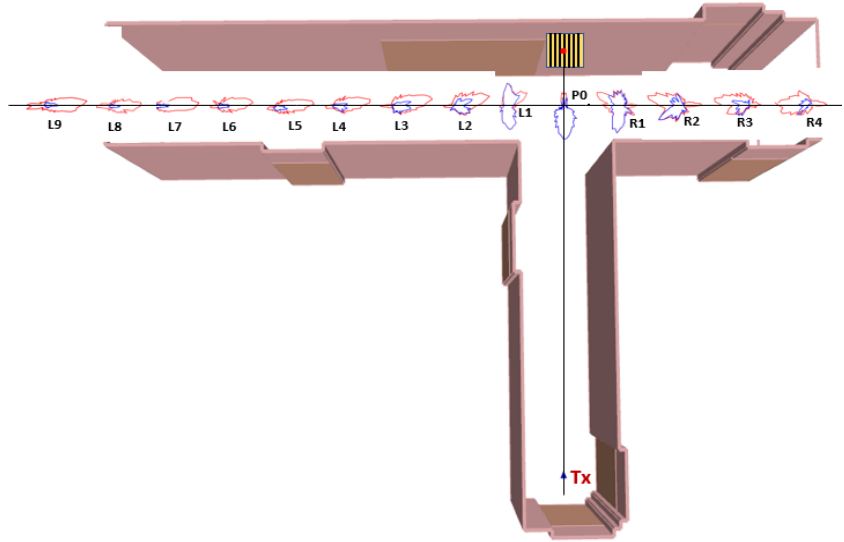

**Fig. 14.** Distribution of the angular power received for each position of Rx.

P0 is the position of Rx in the Tx-reflector alignment, which corresponds to the intersection of the central axes of the two corridors. At position P0, where the two antennas are in LOS, the maximum received power is concentrated in the main lobe and is equal to – 83.8 dBm (Fig. 15). This value is close to that calculated for free space (-82.18 dBm).

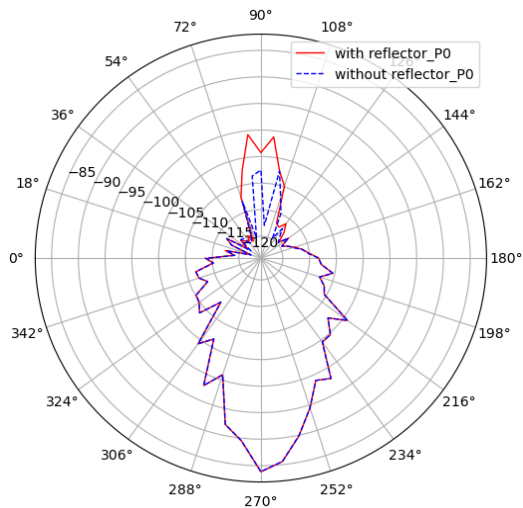

**Fig. 15.** Angular power received for position P0.

The positions R1 to R4 are right-hand Rx positions and L1 to L9 are left-hand positions. For displacements of Rx on positions R1 and L1, the distribution of the received power clearly shows the diffraction effect at the edges of the corridor. Fig. 16 shows in detail the $P_r$ versus the angle of arrival (*AoA*), for positions R1 and L1.

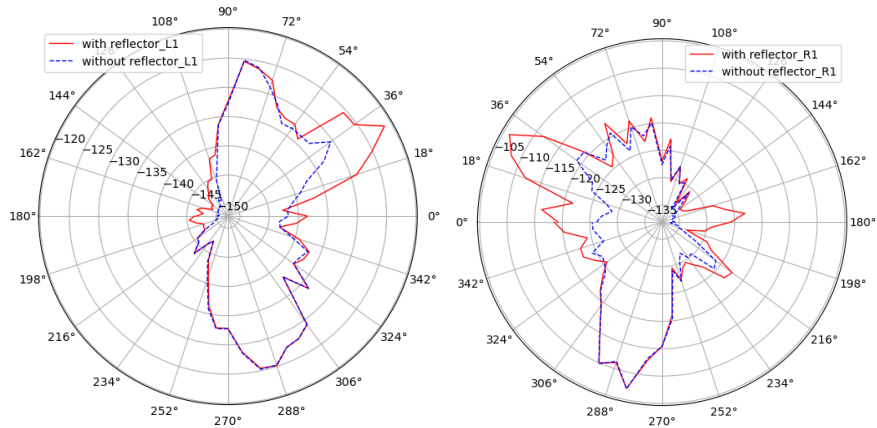

**Fig. 16**. Angular received power for R1 and L1.

These results show that for the most distant positions, the largest $P_r$ is obtained when the Rx antenna is oriented towards the reflector array. Using the reflector increases the received power by approximately 20 dB for position L3. The reflector array provides a significant gain in the link budget. This gain is maintained up to an Rx – P0 distance of 9 m, with an angle of arrival of 6° relative to the central axis of the corridor (Fig. 17).

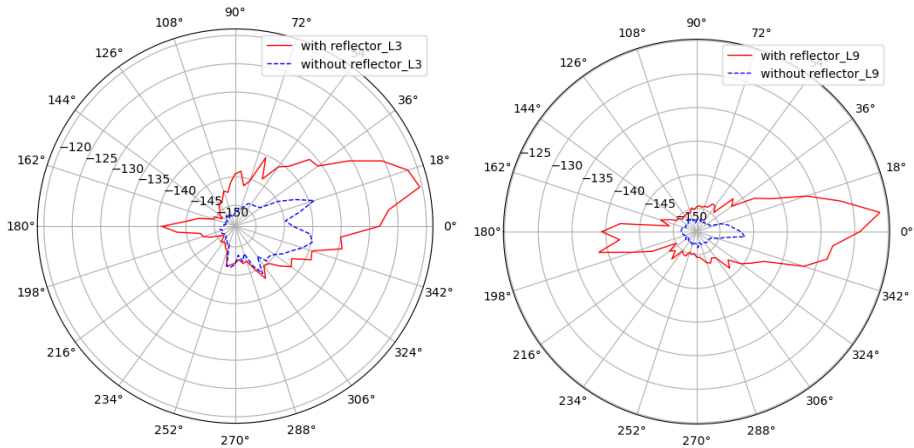

**Fig. 17.** Power received as a function of the angle of arrival at L3 and L9.

These results also show that the maximum received power is practically maintained from position L2 to position L9, i.e. over a distance of 8 m (Fig. 14). Figs. 18 and 19 show the maximum power received ($Max\_P_r$) for the right and left positions. It is clear that the maximum power received is at position P0, which is equal to -83.8 dBm. Fig. 19 clearly shows that the reflector array provides a gain of around 20 dB from position L3.

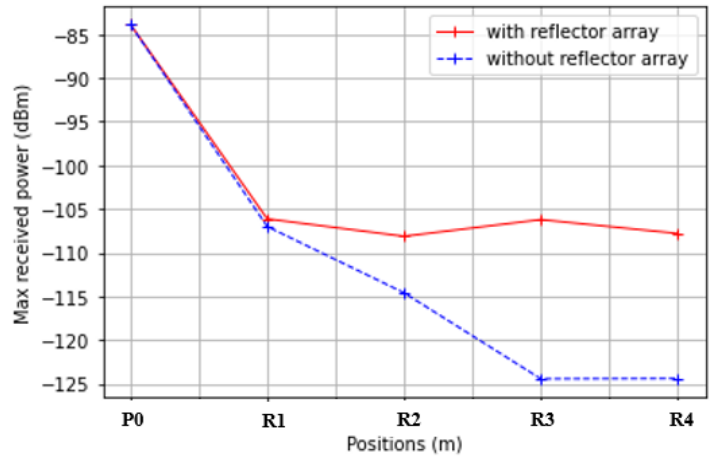

**Fig. 18.** *Max\_P_r* obtained for right-hand positions.

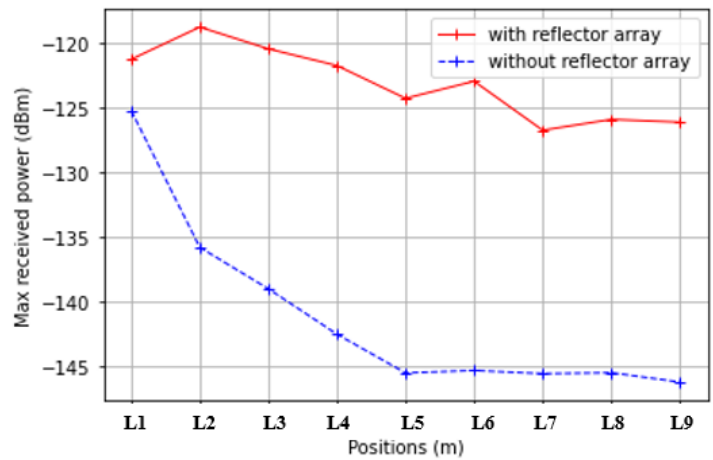

**Fig. 19.** *Max\_P_r* obtained for left-hand positions.

This means that over this distance there is no need to update the beamforming, which can maintain the same dominant trajectory. Finally, this result confirms that the use of a reflector array is an effective solution to improve the radio coverage of indoor mmW off-line-of-sight (NLOS) links.

## 3 Impact of blocking by the human body at 60 GHz

This section presents broadband radio propagation measurements conducted at 60 GHz inside a meeting room. The measurement system used is the same as previously.

### 3.1 Measurement environment

The environment selected for this measurement is a time-invariant (no movement) meeting room whose dimensions are: 6.5 x 2.2 x 2.5 m$^3$. Fig. 20 shows a photo of the measurement environment. In this room, there are two whiteboards with the same dimensions: 2 m x 1.2 m. The door remained closed throughout the measurement campaign.

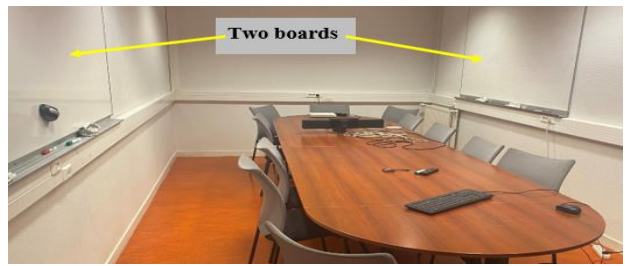

**Fig. 20.** Photo of the measurement environment.

In this room, the presence of the chairs and the table does not modify the measurement results, because of the use of very directional antennas. This room also has a wooden door with the height of 2.3 m and the width of 0.96 m (Fig. 21).

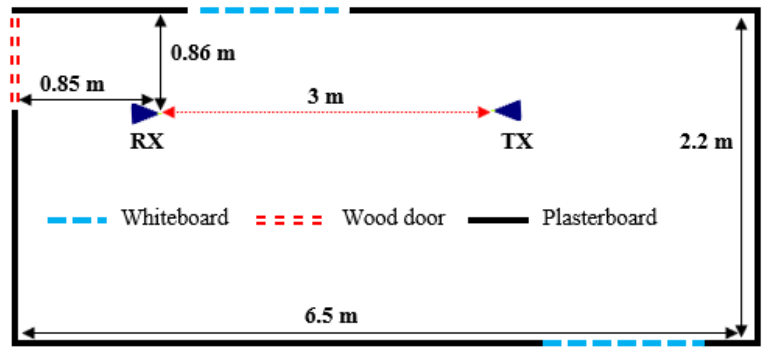

**Fig.21.** Measurement environment in the meeting room.

There was no movement during these measurements. The same horn antennas were used on Tx and Rx sides.

## 3.2 Measurement scenario

The measurements were conducted in a meeting room of the IETR - INSA Rennes. The two antennas were put on supports at the same height of 1.71 m and separated by a distance of 3 m. The height of the two antennas was chosen so that it corresponds to the middle of the whiteboard. The receiving antenna was placed 0.85 m from the door. Its support is fixed on the positioner, which allows the rotation of the antenna. The whiteboard is at a distance of 1.35 m from the door (Fig. 22). The two antennas are equidistant from the center of the board and are placed 0.86 m from the wall of the board. A human blocker, whose width at the shoulders, thickness and height are respectively 45 cm, 13 cm and 1.72 m, was considered.

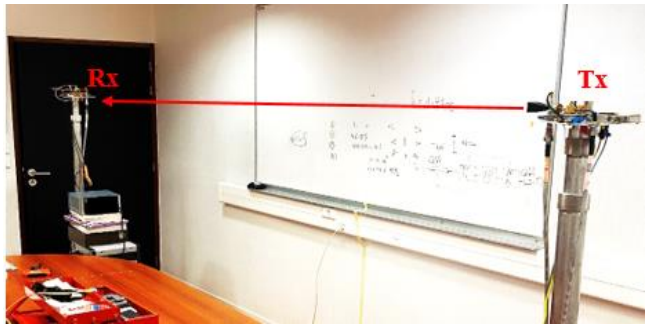

**Fig. 22.** Measurement scenario in the meeting room.

An antenna positioning system was used. Its role is to perform antenna rotations by precise steps for a study of the angle of arrival (*AoA*) of the wave in azimuth. The positioner was used with a mast placed on the motor to support the receiving antenna.

For this measurement campaign, three different scenarios are considered:

- Case where the antennas are in Line of Sight (LOS) and oriented towards each other.

Fig. 23 shows the case where the two antennas are in LOS configuration, oriented towards each other.

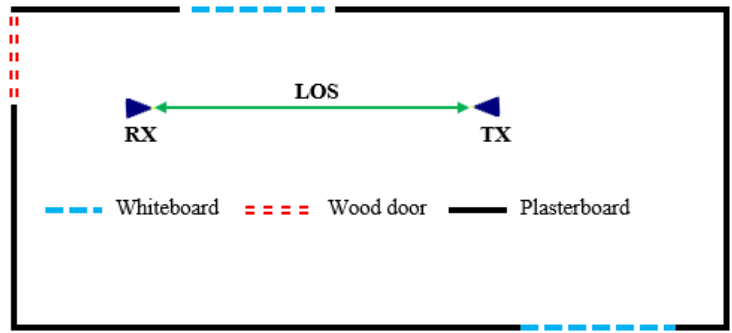

**Fig. 23.** LOS case.

- Case where the direct path is blocked.

In this case, the blocker was placed in the middle of the Tx-Rx distance and oriented towards the receiver, without any movement. The height of the antennas corresponds to the center of the blocker's chest, as shown in Fig. 24. For this, the blocker was placed on a support, so that the center of his chest was at a height of 1.71 m. The comparison with the previous LOS case makes it possible to estimate the attenuation introduced by the human body.

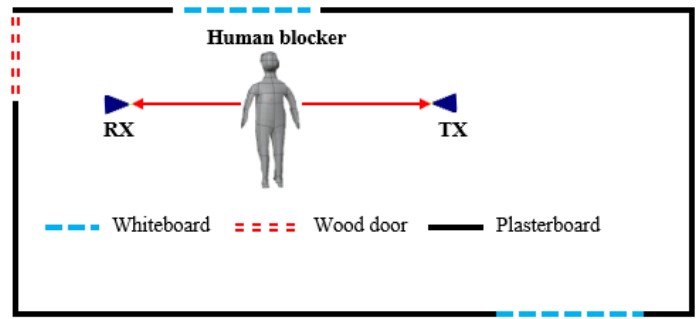

**Fig. 24.** Blocking the direct path.

- Case with blocking the direct path and depointing the antennas by 30° towards the whiteboard.

In this case, the two horn antennas have an azimuth pointing angle of 30° with respect to the Tx-Rx line. With this arrangement, the two antennas were oriented towards the center of the whiteboard (Fig. 25), which allows us to obtain an optimal reflection. In this configuration, the position and height of the blocker and the antennas have not changed.

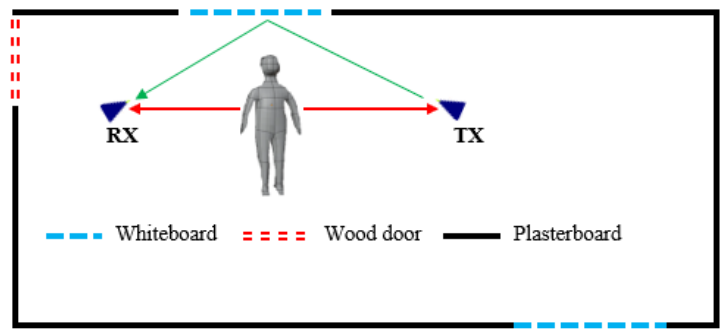

**Fig. 25.** Case with blocking and depointing of the antennas.

Moreover, in these three cases, the receiving antenna rotates 360° in azimuth using the positioner, with a step of 6°. This choice makes it possible to evaluate the channel power delay profile (PDP) on these 60 rotation positions for a study of the paths with the delays and the angles of arrival (*AoA*).

### 3.3 Measurement results

By data post processing, it is possible to represent the power received (in dBm), with and without blocking, for each orientation of the Rx antenna (Fig. 26). We also calculated the impulse response of the propagation channel, in order to study the different paths with their delays.

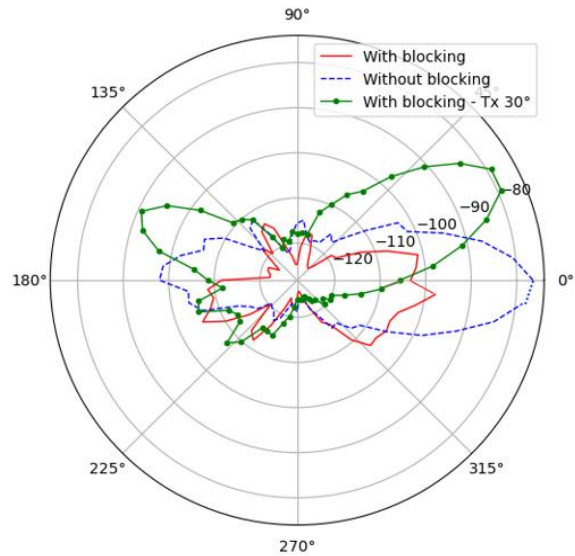

**Fig. 26.** Angular power received (in dBm).

The calculation of the maximum clearance radius of the first Fresnel ellipsoid, at a distance of 1.5 m, gives 0.06 m, which is much lower than the distance that separates the Tx-Rx line from the whiteboard and also from other furniture and equipment in the room. We can therefore consider that we are close to free space conditions, in the case where the two antennas are in direct visibility and without the human blocker. The calculation of the free space *PL* gives 77.54 dB. The *PL* obtained by measurement in the case of direct visibility (i.e. an *AoA* of 0°) and without blocking is 77.2 dB (Fig. 26), which is very close to the value previously calculated in free space condition. With the presence of a human blocker, and taking into account the distances adopted and the beam width of the antennas, it is checked that the main radiation lobe is completely masked by the thorax of the blocker. In this case, when the antennas are pointed towards each other, a path loss of 103 dB is obtained. These results allow us to estimate the losses introduced by the blocker at around 25.8 dB, which is not far from the results presented in [8] at 60 GHz. The path loss is higher and equal to 105.6 dB in the case where only the transmitting antenna is offset by 30° towards the whiteboard and it becomes equal to 79 dB for the same offset of 30° of the two antennas. This shows that with the blocker, depointing the transmit antenna alone results in an increase in path loss of about 2.6 dB and a decrease of 24 dB if both antennas are depointed 30° compared to the blocked path without depointing. So, the reflected path causes a loss of only 1.8 dB compared to the direct path. This low path loss value can be justified by a very small difference in distance between the direct and reflected paths (0.46 m) and by the fact that the reflection is almost-perfect with the board considered to be metallic. Indeed, a calculation of the difference in losses due to the only difference in paths gives 1.25 dB, which makes it possible to estimate the losses by reflection at approximately 0.55 dB.

On the other hand, the losses introduced by human blockage can be estimated by calculating the channel impulse response on a 2 GHz band. This makes it possible to obtain the relative power delay profile (PDP) for the different scenarios considered. Fig. 27 shows the relative PDP, i.e. the PDP normalized to the maximum value computed over all PDPs. The largest value (0 dB) is obtained, as expected, in the case where the two antennas are in LOS (blue curve).

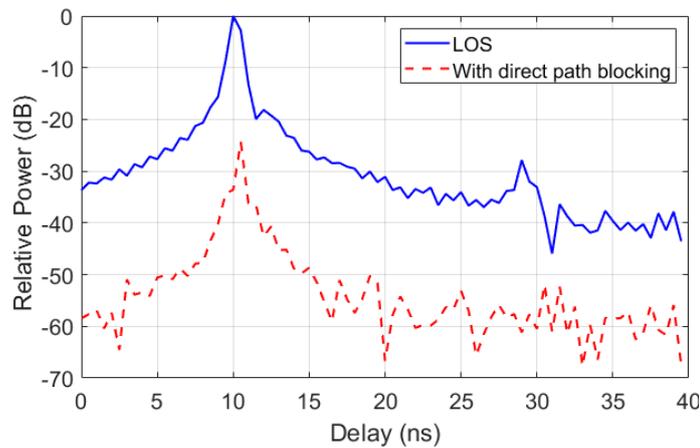

**Fig. 27.** Relative PDP with and without human blocking.

In this case, the direct path arrives with a delay of 10 ns, which corresponds to the Tx-Rx distance of 3 m. With the human blocker, we obtain a relative power value of about - 24.4 dB with a delay of 10.5 ns (red curve). This delay corresponds to a distance of 3.15 m, i.e. a path difference with respect to the direct path of 0.15 m. Thus, the human blocking attenuation of 24.4 dB, obtained by evaluating the relative received power, give a value close to that calculated with the path loss, which is equal to 25.8 dB.

Fig. 30 gives a comparison between the strongest paths in the LOS case and the case where the Tx antenna is oriented 30° towards the middle of the reflective board. This maximum corresponds to the path reflected by the whiteboard. We obtain a relative power value of about - 25.3 dB with a delay of 11.5 ns. Compared with the blocked direct path case, a small relative power increase of 0.8 dB is observed. This can be justified by the fact that we used directional antennas and that the depointing of one of the antennas can allow the signal to bypass the blocker.

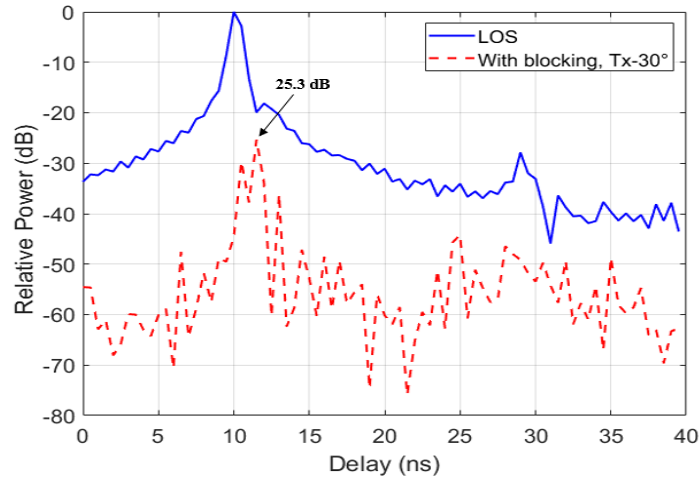

**Fig. 30.** Relative PDP with blocking and Tx offset by 30°.

In the last case, both antennas are pointed towards the center of the whiteboard, with a 30° offset. This configuration seeks to exploit an alternative reflected path that could replace the direct path when the latter is blocked. Fig. 31 shows the relative PDP of this configuration.

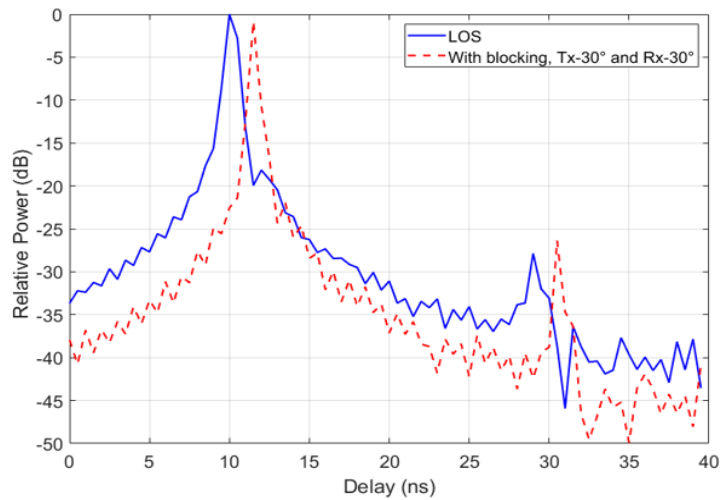

**Fig. 31.** Relative PDP with blocking and depointing of the two antennas by 30°.

In the case where the two antennas are pointed towards the center of the whiteboard, a reflected path is obtained with a delay of 11.5 ns and a relative power of approximately -0.84 dB.
By comparing the received level relative to the direct path without blocking, with that received by the reflected path with blocking and depointing, there is a slight decrease of 0.84 dB for a distance difference of 0.45 m. These results are comparable to those obtained with the path loss calculation, where the reflected path gives additional losses of 1.8 dB compared to the direct path, with a distance difference of 0.46 m. This small difference in path loss can be justified by a very small difference between the lengths of the direct and the reflected paths, and by the fact that the reflection is almost perfect with the whiteboard considered metallic. Therefore, in blocking situations, in order to limit outage times of the radio link, a reflected path can be used as a replacement for the direct path.

## 4 Conclusions

In this chapter, we first studied by measurement the influence of a rectangular metal reflector panel and a reflector antenna array on the radio coverage at 60 GHz in NLOS conditions. Measurements were carried out in an L-shaped corridor with and without the reflector. The results obtained show that with the reflector placed horizontally, then vertically, the received power increases with approximately 24 dB and 12 dB respectively. These results show that the use of the metal panel increases the received power in this NLOS environment.

Then, we studied the use of a reflector antenna array in the T-shaped corridor. These results show that for the most distant positions, the maximum received power is obtained when the Rx horn antenna is oriented towards the direction of the reflector array. The reflector array provides a significant gain in the link budget of around 20 dB for the L3 position in NLOS. These results also show that the same received power and AOA are maintained over a distance of 8 m, so there is no need to update the beamforming. These results confirm that the use of a reflector array is an effective solution to improve radio coverage of indoor mmW out-of-line-of-sight (NLOS) links.

Finally, the results of broadband measurements on the blocking effect by the human body on an indoor radio link at 60 GHz were studied. These results show that the presence of a human blocker introduces additional losses, generally between 24 and 26 dB. On the other hand, the calculation of the channel impulse response makes it possible to highlight the multipath phenomenon and leads to obtaining the PDPs for the different measurement configurations. In particular, in the presence of a reflection on a metallic surface, the reflected path can be used when the direct path is blocked, using both transmit and receive beamforming.

However, in order to achieve a more precise modeling of the mmW propagation channel, further measurement campaigns remain necessary to be performed in other environments with less favorable reflection conditions, and using other types of antennas and reflective panels. Other parameters can also be studied such as the location of the blocker, the distance between the antennas or between the Tx-Rx line and the reflector and the height of the antennas.

## Acknowledgment

This work was financially supported by the French National Research Agency as part of the ANR MESANGES project (ANR-20-CE25-0016).